\ifx\synctex\undefined\else\synctex=1\fi
\pdfoutput=1
\documentclass[submission]{eptcs}

\providecommand{\event}{HCVS 2016}

\usepackage{graphicx}

\usepackage{amsmath} 
\usepackage[linesnumbered,boxruled,vlined,english]{algorithm2e}
\usepackage{alltt}
\usepackage{tikz}
\usepackage{adjustbox} 

\usepackage{amssymb}
\usepackage{upquote}
\usepackage{fancyvrb} 

\usepackage{float}
\usepackage{wrapfig}

\usepackage{hyperref}

\usepackage{commenting}

\declareauthor{jg}{John}{blue}
\declareauthor{pg}{Pierre}{orange!50}
\declareauthor{bk}{Bish}{red}

\title{ Solving non-linear Horn clauses using a linear Horn clause solver \thanks{The research leading to these results
   has been supported by EU
  FP7 project 318337, \emph{ENTRA - Whole-Systems Energy Transparency},   EU FP7 project 611004, \emph{coordination and support action ICT-Energy}, 
   EU FP7 project 610686, \emph{POLCA - Programming Large Scale Heterogeneous Infrastructures},  Madrid Regional Government project S2013/ICE-2731, \emph{N-Greens Software - Next-GeneRation Energy-EfficieNt Secure Software}, and the Spanish Ministry of Economy and
Competitiveness project No.\@ TIN2015-71819-P, \emph{RISCO - RIgorous analysis of Sophisticated COncurrent and distributed systems}.}}

\author{Bishoksan Kafle \institute{Roskilde University, Denmark} \email{kafle@ruc.dk}
\and
John P. Gallagher \institute{Roskilde University, Denmark} 
\institute{IMDEA Software Institute, Spain}
 \email{jpg@ruc.dk}
\and 
 Pierre  Ganty \institute{IMDEA Software Institute, Spain} \email{pierre.ganty@imdea.org}
 }

\def\titlerunning{Solving non-linear Horn clauses using a linear Horn clause solver}
\def\authorrunning{ Kafle, Gallagher and Ganty}

\begin{document}
\maketitle

\newcommand{\integ}{{\sf int}}
\newcommand{\listint}{{\sf listint}}
\newcommand{\other}{{\sf other}}
\newcommand{\true}{\textsc{True}}
\newcommand{\false}{\textsc{False}}
\newcommand{\Bin}{{\sf Bin}}
\newcommand{\Dep}{{\sf Dep}}
\newcommand{\g}{{\sf g}}
\newcommand{\nong}{{\sf ng}}
\newcommand{\OL}{{\cal O}}
\newcommand{\M}{{\sf M}}
\newcommand{\R}{{\cal R}}
\newcommand{\A}{\mathcal{A}}

\newcommand{\body}{\mathcal{B}}
\newcommand{\B}{{\cal B}}
\newcommand{\C}{{\cal C}}
\newcommand{\D}{{\cal D}}
\newcommand{\X}{{\cal X}}
\newcommand{\V}{{\cal V}}
\newcommand{\Q}{{\cal Q}}
\newcommand{\F}{{\sf F}}
\newcommand{\N}{{\cal N}}
\newcommand{\Lang}{{\cal L}}
\newcommand{\powerset}{{\cal P}}
\newcommand{\FTA}{{\cal FT\!A}}
\newcommand{\Term}{{\sf Term}}
\newcommand{\Empty}{{\sf empty}}
\newcommand{\nonEmpty}{{\sf nonempty}}
\newcommand{\compl}{{\sf complement}}
\newcommand{\args}{{\sf args}}
\newcommand{\preds}{{\sf preds}}
\newcommand{\gnd}{{\sf gnd}}
\newcommand{\lfp}{{\sf lfp}}
\newcommand{\psharp}{P^{\sharp}}
\newcommand{\minimize}{{\sf minimize}}
\newcommand{\headterms}{\mathsf{headterms}}
\newcommand{\solvebody}{\mathsf{solvebody}}
\newcommand{\solve}{\mathsf{solve}}
\newcommand{\fail}{\mathsf{fail}}
\newcommand{\member}{\mathsf{memb}}
\newcommand{\ground}{\mathsf{ground}}

\newcommand{\raf}{{\sf raf}}
\newcommand{\qa}{{\sf qa}}
\newcommand{\spl}{{\sf split}}

\newcommand{\transitions}{\mathsf{transitions}}
\newcommand{\nonempty}{\mathsf{nonempty}}
\newcommand{\dom}{\mathsf{dom}}

\newcommand{\Args}{\mathsf{Args}}
\newcommand{\id}{\mathsf{id}}
\newcommand{\type}{\tau}
\newcommand{\restrict}{\mathsf{restrict}}
\newcommand{\any}{\top}
\newcommand{\dyn}{\top}
\newcommand{\dettypes}{{\sf dettypes}}
\newcommand{\Atom}{{\sf Atom}}

\newcommand{\chc}{{\sf chc}}
\newcommand{\deriv}{{\sf deriv}}

\newcommand{\vars}{\mathsf{vars}}
\newcommand{\Vars}{\mathsf{Vars}}
\newcommand{\range}{\mathsf{range}}
\newcommand{\varpos}{\mathsf{varpos}}
\newcommand{\varid}{\mathsf{varid}}
\newcommand{\argpos}{\mathsf{argpos}}
\newcommand{\elim}{\mathsf{elim}}
\newcommand{\pred}{\mathsf{pred}}
\newcommand{\predfuncs}{\mathsf{predfuncs}}
\newcommand{\project}{\mathsf{project}}
\newcommand{\reduce}{\mathsf{reduce}}
\newcommand{\positions}{\mathsf{positions}}
\newcommand{\contained}{\preceq}
\newcommand{\equivalent}{\cong}
\newcommand{\unify}{{\it unify}}
\newcommand{\Iff}{{\rm iff}}
\newcommand{\Where}{{\rm where}}
\newcommand{\State}{\mathsf{S}}
\newcommand{\qmap}{{\sf qmap}}
\newcommand{\fmap}{{\sf fmap}}
\newcommand{\ftable}{{\sf ftable}}
\newcommand{\Qmap}{{\sf Qmap}}
\newcommand{\states}{{\sf states}}
\newcommand{\head}{\tau}
\newcommand{\atomconstraints}{\mathsf{atomconstraints}}
\newcommand{\thresholds}{\mathsf{thresholds}}
\newcommand{\term}{\mathsf{Term}}
\newcommand{\trees}{\mathsf{trees}}
\newcommand{\renames}{\rho}
\newcommand{\renameps}{\rho_2}
\newcommand{\predicates}{\mathsf{Predicates}}
\newcommand{\query}{\mathsf{q}}
\newcommand{\ans}{\mathsf{a}}
\newcommand{\trace}{\mathsf{tr}}
\newcommand{\constr}{\mathsf{constr}}
\newcommand{\Iproj}{\mathsf{proj}}
\newcommand{\SAT}{\mathsf{SAT}}
\newcommand{\interpolant}{\mathsf{interpolant}}
\newcommand{\unknown}{?}
\newcommand{\rhs}{{\sf rhs}}
\newcommand{\lhs}{{\sf lhs}}
\newcommand{\unfold}{{\sf unfold}}
\newcommand{\arity}{{\sf ar}}
\newcommand{\AND}{{\sf AND}}

\newcommand{\atmost}[1]{\le #1}
\newcommand{\exactly}[1]{=#1}
\newcommand{\exceeds}[1]{>#1}
\newcommand{\anydim}[1]{\ge 0}

\def\ll{[\![}
\def\rr{]\!]}

\newcommand{\sset}[2]{\left\{~#1  \left|
                               \begin{array}{l}#2\end{array}
                          \right.     \right\}}

\newcommand{\qin}{\hspace*{0.15in}}
\newenvironment{SProg}
     {\begin{small}\begin{tt}\begin{tabular}[t]{l}}%
     {\end{tabular}\end{tt}\end{small}}
\def\anno#1{{\ooalign{\hfil\raise.07ex\hbox{\small{\rm #1}}\hfil%
        \crcr\mathhexbox20D}}}

\newtheorem{definition}{Definition}
\newtheorem{example}{Example}
\newtheorem{corollary}{Corollary}

\newtheorem{lemma}{Lemma}
\newtheorem{theorem}{Theorem}
\newtheorem{proposition}{Proposition}
\newtheorem{property}{Property}

\begin{abstract}

In this paper we show that checking satisfiability of
a set of non-linear Horn clauses (also called a non-linear Horn clause program)
can be achieved using a solver for linear Horn clauses.  
We achieve this by interleaving a  \emph{program transformation} with a satisfiability checker for linear Horn clauses (also called a \emph{solver} for linear Horn clauses). The program transformation is based on  the notion of \emph{tree dimension}, which we apply to a set of non-linear clauses, yielding a set whose derivation trees have bounded dimension. Such a set of clauses can be linearised. The main algorithm then proceeds by applying the linearisation transformation and solver for linear Horn clauses to a sequence of sets of clauses with successively increasing dimension bound.
The approach is then further developed by using a solution of clauses of lower dimension to (partially) linearise clauses of higher dimension.
 We constructed a prototype implementation of this approach and performed some experiments on a set of verification problems, which shows some promise.

\end{abstract}

\section{Introduction}
Many software verification problems can be reduced to checking satisfiability of a set of Horn clauses (the \emph{verification conditions}).
In this paper we propose an 
approach for checking satisfiability of a set of non-linear Horn clauses (clauses whose body contains more than one non-constraint atom) using a linear Horn clause solver.  
A  \emph{program transformation} based on the notion of \emph{tree dimension}  is applied to a set of non-linear Horn clauses;  this gives a set of clauses that can be linearised and then solved using a \emph{linear  solver} for Horn clauses.  This combination of dimension-bounding, linearisation and then solving with a linear solver is repeated for successively increasing dimension.  The dimension of a tree is a measure of its non-linearity -- for example a linear tree  (whose nodes have at most one child) has dimension zero while a complete binary tree has dimension equal to its height. 

 A given set of Horn clauses $P$ can be transformed into a new set of clauses
$P^{\atmost{k}}$, whose derivation trees are the subset of $P$'s derivation trees with dimension at most $k$.  
It is known that $P^{\atmost{k}}$ can be transformed to a linear set of clauses preserving satisfiability;  hence if we can find a model of the linear set of clauses then the original clauses $P^{\atmost{k}}$ also have a model.

The algorithm terminates with success if a model (solution) $M$ of $P^{\atmost{k}}$  is also a model (after appropriate translation of predicate names)  of $P$.   However if $M$ is not a solution of $P$, then we proceed to generate $P^{\atmost{k+1}}$ and repeat the procedure. The algorithm terminates  if $P^{\atmost{k}}$ is shown to be unsatisfiable (unsafe) for some $k$, since this implies that $P$ is also unsatisfiable. 

A more sophisticated version of the algorithm attempts to use the model $M$ of $P^{\atmost{k}}$ to (partially) linearise $P^{\atmost{k+1}}$. We can exploit the model of $P^{\atmost{k}}$ in the following way; if $P^{\atmost{k+1}}$ has a counterexample that does not use 
the (approximate) solution $M$ for $P^{\atmost{k}}$ , then $P$ is unsatisfiable. We continue this process successively for increasing value of $k$ until we find a solution or a counterexample  to $P$,  or until resources are exhausted.

As an example program, we consider a set of constrained Horn clauses $P$ in Figure  \ref{exprogram} which defines the Fibonacci function. It is an interesting problem since its derivations are trees whose dimensions depend on an input argument.  The last clause represents a property of the Fibonacci function expressed as an integrity constraint.

 \begin{figure}[t]
\centering
\begin{BVerbatim}
c1. fib(A, B):- A>=0,  A=<1, B=A.
c2. fib(A, B) :- A > 1, A2 = A - 2, fib(A2, B2),
           A1 = A - 1, fib(A1, B1), B = B1 + B2.
c3. false:- A>5, fib(A,B), B<A.          
\end{BVerbatim}
\caption{Example CHCs Fib defining a Fibonacci function.}
 \label{exprogram}

\end{figure}

We have made a prototype implementation of this approach and performed some experiments on a set of software verification problems, which shows some promise. The main contributions of this paper are as follows.

\begin{enumerate}
\item We present a linearisation procedure for dimension-bounded Horn clauses using partial evaluation (Section \ref{linearisation}).
\item We give an 
algorithm for solving a set of non-linear Horn clauses using a linear Horn clause solver (Section \ref{procverfication}).
\item  We demonstrate the feasibility of our approach in practice  applying it to non-linear Horn clause  problems (Section \ref{experiments}).

\end{enumerate}




\section{Preliminaries}
\label{prelim}


A constrained Horn clause (CHC)  is a first order  formula of the form 
$ p(X) \leftarrow   \C , p_1(X_1) , \ldots , p_k(X_k) $ ($k \ge 0$) (using Constraint Logic Programming syntax),  where $\C$ is a conjunction of \emph{constraints} with respect to some constraint theory, $X_i, X$  are (possibly empty) vectors of distinct \emph{variables}, $p_1,\ldots,p_k, p$ are \emph{predicate symbols}, $p(X)$ is the \emph{head} of the clause and $\C , p_1(X_1) , \ldots , p_k(X_k)$ is the \emph{body}.  
An atomic formula, or simply \emph{atom}, is a formula $p(t)$ where $p$ is a non-constraint predicate symbol and $t$ a tuple of arguments. Atoms are sometimes written as $A$, $B$ or $H$, possibly with sub- or superscripts.

A clause is called \emph{non-linear} if it contains more than one atom in the body, otherwise it is called \emph{linear}.  A set of Horn clauses $P$ is called  linear if $P$ only contains linear clauses, otherwise it is called non-linear.  \emph{Integrity constraints} are a special kind of Horn clauses whose head is  $\mathit{false}$  where $\mathit{false}$ is always interpreted as $\false$. A set of Horn clauses is sometimes called a \emph{(constraint logic) program}.

An \emph{interpretation} of a set of CHCs is represented as a set of \emph{constrained facts} of the form $A \leftarrow \C$ where $A$ is an atomic formula $p(Z)$ where $Z$ is a tuple of distinct variables and $\C$ is a constraint over $Z$ with respect to some constraint theory.  An interpretation that makes each clause in $P$ $\true$ is called a \emph{model} of $P$. We say a set of Horn clause $P$ (including integrity constraints) is safe (solvable) if{}f it has a model.
 In some works e.g. \cite{DBLP:conf/sas/BjornerMR13,McmillanR2013},  a model is also called a \emph{solution} and we use them interchangeably in this paper.

A  labeled tree $c(t_1,\ldots,t_k)$ ($k \ge 0$) is a tree whose nodes are labeled by identifiers, where $c$ is the label of the root and $t_1,\ldots,t_k$ are labeled trees, the children of the root.   

\begin{definition}[Tree dimension (adapted from \cite{DBLP:conf/stacs/EsparzaKL07})]\label{treedim}
  Given a  labeled tree $t= c(t_1,\ldots,t_k)$, the tree dimension of $t$ represented as \(\mathit{dim}(t)\) is defined as follows: 

  \[
  \mathit{dim}(t)= \begin{cases}
  0 & \text{if } k=0 \\ 
  \max_{ i \in [1..k]} \mathit{dim}(t_i) &\text{if  there  is  a  unique  maximum}\\ 
  \max_{ i \in [1..k]} \mathit{dim}(t_i)+1 &\text{otherwise } 
\end{cases}
  \]

\end{definition}
Given a set of Horn clauses, we associate with each clause $ p(X) \leftarrow   \C , p_1(X_1) , \ldots , p_k(X_k) $ a unique identifier $c$ whose \emph{arity} is $k$.   

Labelled trees can represent Horn clause derivations, where node labels are clause identifiers.
\begin{definition}[Trace tree]\label{def:tracetree}
A \emph{trace tree} for an atom $A$ in a set of Horn clauses $P$ is a labelled tree $c(t_1,\ldots,t_k)$ if $c$ is a clause identifier for a clause $A \leftarrow   \C, A_1, \ldots, A_k$ in $P$ (with variables suitably renamed) and $t_1,\ldots,t_k$ are trace trees for $A_1, \ldots, A_k$ in $P$ respectively.  
\end{definition}

There is a one-one correspondence between  \emph{trace trees} and derivation trees of Horn clauses up to variable renaming. Thus when we speak about the dimension of a Horn clause derivation, we refer to the dimension of its corresponding  \emph{trace tree}. 

\begin{figure}[h!]
  \centering
    \includegraphics[width=0.70 \textwidth, height=45mm]{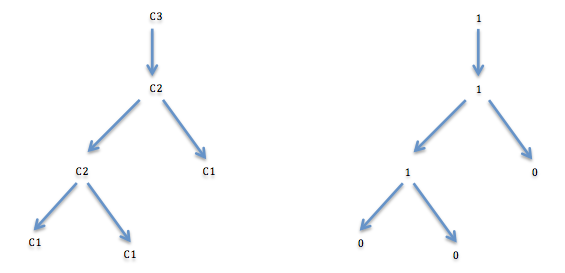}
    \caption{(a) a \emph{trace tree}   and (b) its tree dimension.\label{fig:treedimension}}
\end{figure}

Using the clauses shown in Figure  \ref{exprogram} along with their  identifiers, Figure~\ref{fig:treedimension} (a) shows a  \emph{trace tree} $t=c_3(c_2(c_2(c_1,c_1), c_1))$  and  Figure~\ref{fig:treedimension} (b) shows  its tree dimension.  It can be seen that $\mathit{dim}(t)=1$.  

 To make the paper self contained, we describe the transformation to produce a dimension-bounded set of clauses. Given a set of CHCs $P$ and  $k \in \mathbb{N}$, we split each predicate $p$ occurring in $P$ into the predicates $p^ {\atmost{d}}$ and $p^ {\exactly{d}}$ where $d \in \{ 0,1,\ldots,k\}$. An atom with predicate $p^ {\atmost{d}}$ or $p^ {\exactly{d}}$ is denoted $H^ {\atmost{d}}$ or $ H^ {\exactly{d}}$ respectively.  Such atoms have derivation trees of dimension at most $d$ and exactly $d$ respectively. 

\begin{definition}[ At-most-\(k\)-dimension program $P^{\atmost{k}}$]
\label{kdim-most}
Let $P$ be a set of CHCs. $P^{\atmost{k}}$ consists of the following clauses (adapted  from \cite{DBLP:journals/iandc/LuttenbergerS16, DBLP:journals/corr/KafleGG15}):

\begin{enumerate}

\item Linear clauses:

 If $H  \leftarrow  \C \in P$ , then $H^ {\exactly{0}} \leftarrow  \C \in P^{\atmost{k}}$.
 
 If $H  \leftarrow  \C, B_1  \in P$  then $H^ {\exactly{d}} \leftarrow   \C,  B_1 ^{\exactly{d}} \in P^{\atmost{k}}$ for  $0 \le d \le k$.

\item Non-linear clauses: 

 If $H  \leftarrow  \C, B_1 , B_2 , \ldots,B_r  \in P$ with $r>1$ and one of the following holds:
\begin{itemize}
\item For $1 \le d \le k$, and $1 \le j \le r$:

Set  $Z_j =B_j^{\exactly{d}}$ and $Z_i = B_i ^{\atmost{d-1}}$ for $1 \le i \le r \wedge i \neq j$. Then: 
$H^ {\exactly{d}} \leftarrow   \C, Z_1,\ldots, Z_r \in P^{\atmost{k}}$.

\item 
 For  $1 \le d \le k$, and $J \subseteq \{1,\ldots, r\}$ with $\vert J \vert = 2$:
 
Set $Z_i =B_i^ {\exactly{d-1}}$ if $i \in J$ and $Z_i =B_i^ {\atmost{d-1}}$ if $i \in \{1,\ldots,r\} \setminus J$. 
If all \(Z_i\) are defined, i.e., \(d\geq 2\) if \(r>2\), then:
$H^ {\exactly{d}} \leftarrow   \C, Z_1, \ldots, Z_{r} \in P^{\atmost{k}}$.

\end{itemize}

\item $\epsilon$-clauses: 

$H^ {\atmost{d}} \leftarrow H^ {\exactly{e}} \in P^{\atmost{k}}$ for $0 \le d \le k$ , and every $0 \le e \le d \leq k$.

\end{enumerate}
\end{definition}

\noindent
$P^{\atmost{k}}$ is also called the \(k\)-dimension-bounded program corresponding to $P$. When the value of $k$ is not important, any program generated using the Definition \ref{kdim-most} is called a dimension-bounded program. The relation between $P$ and its $k$-dimensional program is given by in the Proposition \ref{underapproximation} where  $\models$ is the usual ``logical consequence'' operator. 

\begin{proposition}[Relation between $P$ and $P^{\atmost{k}}$]
\label{underapproximation}
Let $P$ be a program and $P^{\atmost{k}}$ ($k\geq0$) be its $k$-dimension-bounded program. Let $p(t)$ be an atom where $p$ is a predicate of $P$ and $p^ \star(t)$ ($ \star \in \{=, \leq\}$) be an atom where $p^{\star}$ is a predicate of $P^{\atmost{k}}$. Then we have: $P^{\atmost{k}} \models p^ \star(t) \implies  P \models p(t)$. 
\end{proposition}

In other words,  Proposition \ref{underapproximation} says that the set of facts that can be derived from $P^{\atmost{k}}$ is a subset of the set of facts that can be derived from $P$, taking the predicate renaming into account. In this sense $P^{\atmost{k}}$ is an \emph{under-approximation} of $P$. In particular, if $P^{\atmost{k}} \models \mathtt{false}^ \star$ then $P \models \mathtt{false}$.

Let $S$ be an interpretation of a dimension-bounded set of clauses $P^{\atmost{k}}$.  That is, $S$ is a set of constrained facts
of the form $H^ {\exactly{d}} \leftarrow   \C$ or $H^ {\atmost{d}} \leftarrow \C$.  An interpretation of $P$ is constructed from $S$ as follows.

\begin{definition}[$S^{\uparrow P^{\atmost{k}}}$: an  interpretation of $P$ constructed from an interpretation of $P^{\atmost{k}}$]
\label{liftingDefinition}
Let $S$ be an interpretation of $P^{\atmost{k}}$. Then $S^{\uparrow P^{\atmost{k}}}$ is the following set of constrained facts.
 
$$S^{\uparrow P^{\atmost{k}}} =   \{ p(X) \leftarrow \bigvee \{\C \mid p^ {\exactly{d}} (X)\leftarrow   \C \in S \vee p^ {\atmost{d}}(X) \leftarrow \C \in S\} 
			\mid ~ d \in \{0...k\} ~ and ~ p \mathrm{~is~a~predicate~in~} P\}$$
The set $S^{\uparrow P^{\atmost{k}}}$ is a disjunctive interpretation of $P$ where the interpretation of $p$ is the disjunction of the interpretations of the corresponding dimension-bounded versions of $p$ in  $P^{\atmost{k}}$.

\end{definition}

 \begin{figure}[t]
 \centering
\begin{BVerbatim}
	%linear clauses
        1. fib(0)(A,B) :- A>=0, A=<1, B=A.
        2. false(0) :- A>5, B<A, fib(0)(A,B).
        %epsilon-clauses
        3. false[0] :- false(0).
        4. fib[0](A,B) :- fib(0)(A,B).
\end{BVerbatim}
\caption{Fib$^ {\atmost{0}}:$ at-most-\(0\)-dimension program of {Fib}.}
 \label{fib0dim}
\end{figure}

 \begin{figure}[t]
 \centering
\begin{BVerbatim}
fib(0)(A,B) :- B=A,  A=<1, A>=0.
fib(1)(A,B) :-  B=F+D, C=A-2,
            E=A-1,  A>1, fib[0](E,F), fib(1)(C,D).
fib(1)(A,B) :-  B=F+D, C=A-2, E=A-1,
            A>1, fib[0](C,D), fib(1)(E,F).
fib(1)(A,B) :-  B=F+D,  C=A-2,  E=A-1,
            A>1, fib(0)(C,D), fib(0)(E,F).
false(1) :-  B<A,  A>5,  fib(1)(A,B).
false(0) :-  B<A,  A>5, fib(0)(A,B).
false[1] :-  false(1).
false[1] :- false(0).
false[0] :-  false(0).
fib[1](A,B) :-  fib(1)(A,B).
fib[1](A,B) :- fib(0)(A,B).
fib[0](A,B) :- fib(0)(A,B).
\end{BVerbatim}
\caption{Fib$^ {\atmost{1}}:$ at-most-\(1\)-dimension program of {Fib}.}
 \label{fib1dim}
\end{figure}

The  at-most-\(0\)-dimension program  of  {Fib} in Figure \ref{exprogram} is depicted in Figure \ref{fib0dim}.  In textual form we represent a predicate $p^{\atmost{k}}$ by \texttt{p[k]} and a predicate $p^{\exactly{k}}$ by \texttt{p(k)}. The  at-most-\(1\)-dimension program  of  {Fib} in Figure \ref{exprogram} is depicted in Figure \ref{fib1dim}. 
Note that  \(0\)-dimension program is included in \(1\)-dimension program. 
In general,  all the clauses in $P^{\atmost{k}}$ are also in $P^{\atmost{k+1}}$. This provides a basis for an iterative strategy for a bounded set of Horn clauses. Since some programs have derivation trees of unbounded dimension, trying to verify a property for its increasing dimension separately is not a practical strategy. It only becomes a viable approach  if a solution of $p^{\atmost{k}}$ for some $k\geq 0$ is general enough to hold for  all dimensions of $P$.

\section{Linearisation strategies for dimension-bounded set of Horn clauses}
\label{linearisation}

In this section, we present linearisation strategies for set of clauses of bounded dimension.  It is known \cite{DBLP:journals/tcs/AfratiGT03} that a dimension-bounded set of clauses can be linearised, preserving satisfiability.  In this section we describe a practical technique for linearisation, based on partial evaluation of an interpreter.

\subsection{Linearisation based on partial evaluation}
\label{pe}

\emph{Partial evaluation} (PE)
 has been studied for a variety of languages including logic programs \cite{Jones-Gomard-Sestoft,DBLP:conf/pepm/Gallagher93,DBLP:conf/lopstr/Leuschel94,DBLP:journals/scp/Jones04,DBLP:journals/iandc/LeuschelV14}. We follow the pattern of transforming a program (a set of Horn clauses) by specialising an interpreter for that program \cite{Gallagher-86,DBLP:journals/scp/Jones04}.  Let $PE$ be a partial evaluator, $I$ an interpreter and $P$ an object program.  Then the partial evaluation of $I$ with respect to $P$, denoted $PE(I,P)$, represents the ``compilation" of $P$ using the semantics given by $I$.

We first write an interpreter for Horn clause programs, which is also written as a set of Horn clauses. Given a (possibly empty) conjunction of atoms (called a \emph{goal}) the interpreter constructs a derivation, implementing a standard left-to-right, depth-first search. In the interpreter predicate \texttt{solve(Gs)},  \texttt{Gs} is the goal, represented as a list of atoms.  The basic step of the interpreter is represented by the clauses for \texttt{solve(Gs)} shown in Figure \ref{fig:interpreter1}. If the conjunction is not empty, its first atom \texttt{G} is selected along with a matching Horn clause $G \leftarrow Cs,B$ in the program being interpreted, where $Cs$ is a conjunction of constraints and $B$ is a conjunction of atoms. This clause is represented by \texttt{hornClause(G,Cs,B)} in the interpreter.  The body of the clause is conjoined with the remaining goal atoms, and the derivation continues with the new goal \texttt{Gs1}.  If the conjunction is empty, the derivation is successful (second clause).

\begin{figure}[h!]
\centering
\begin{BVerbatim}
solve([G|Gs]) :-
   hornClause(G,Cs,B), solveConstraints(Cs), append(B,Gs,Gs1),
   solve(Gs1).
solve([]).
\end{BVerbatim}
\caption{Depth-first interpreter for Horn clauses}
\label{fig:interpreter1}
\end{figure}
To interpret a dimension-bounded set of clauses (say the bound is \(k\)), we use the fact that in all successful runs of the interpreter in which goals are selected in increasing order of dimension, the size of the conjunction of goals (that is, the length of the argument of \texttt{solve}) has an upper bound related to $k$. This bound is known as the \emph{index} of the set of clauses and is given as $(i-1)*k+1$, where $i$ is the maximum number of non-constraint atoms in the body of clauses \cite{DBLP:journals/ipl/EsparzaGKL11}. Given this index, we can augment the interpreter with a check on the size of the conjunction, ensuring that it never exceeds the index.  In addition, due to the requirement of increasing dimension in the selection of atoms, a left-to-right computation rule is not sufficient; therefore we permute the set of atoms in each clause body, since in at least one permutation the goals will be ordered by dimension. With these changes the interpreter remains complete for clauses of the given maximum index, at the possible cost of some redundancy in the search. 

These additions result in the interpreter whose top level is shown in Figure \ref{fig:interpreter}.  Let the interpreter predicate \texttt{solve(Gs,Index,L)} mean that the conjunction of goals \texttt{Gs} is to be solved, and \texttt{L}, \texttt{Index} are numbers representing the size of \texttt{Gs} and the maximum size of the stack of goals.

\begin{figure}[h!]
\centering
\begin{BVerbatim}
go(Index) :-
   solve([false],Index,1).
solve([G|Gs],Index,L) :-
   hornClause(G,Cs,B), solveConstraints(Cs),
   length(B,L1), L2 is L1+L-1, L2 =< Index,
   perm(B,B1), append(B1,Gs,Gs1),
   solve(Gs1,Index,L2).
solve([],_,_).

\end{BVerbatim}
\caption{Interpreter for linearisation}
\label{fig:interpreter}
\end{figure}

\paragraph{Partial evaluation of the interpreter.}
Given a set of facts of the form \texttt{hornClause(G,Cs,B)} representing the Horn clauses to be linearised, and some value of \texttt{Index}, the interpreter can be partially evaluated.  We use Logen \cite{DBLP:conf/pepm/LeuschelEVCF06} to perform the partial evaluation with respect to a call to \texttt{go(Index)}, which initiates a proof of the goal \texttt{false} (see first clause of interpreter).  All interpreter computations are partially evaluated except for the calls to \texttt{solve(Gs,Index,L)} and the execution of constraints within the goal \texttt{solveConstraints(Cs)}.  Furthermore Logen performs standard structure-flattening and predicate renaming operations, yielding a set of clauses of the form \texttt{solve\_i(X) :- Cs, solve\_j(Y)}, where \texttt{solve\_i(X)} and \texttt{solve\_j(Y)} are instantiations of \texttt{solve(Gs,Index,L)} and \texttt{Cs} is a constraint. Thus the resulting clauses are linear, and furthermore preserve the meaning of the original clauses as given by the interpreter, by correctness of the partial evaluation procedure. The linearisation procedure is independent of the constraint theory underlying the clauses.

\begin{proposition}\label{pe-prop}
  Let $P$ be a program and $P^{\atmost{k}}$ ($k\geq 0$) be its $k$-dimension-bounded program.  Let $i$ be the maximum number of atoms in clause bodies of $P$.  Let $\mathtt{Index} = (i-1)*k+1$.   Let $P'$ be a partial evaluation of the interpreter in Figure \ref{fig:interpreter}, with respect to  $P$ and the goal $\mathtt{go(Index)}$.  Then $P^{\atmost{k}} \models \mathit{false^{\atmost{k}}}$ if{}f $P' \models \mathtt{go(Index)}$.  
\end{proposition}
Furthermore $P'$ is linear if the partial evaluator follows the strategy described above.
Combining Propositions \ref{pe-prop} and \ref{underapproximation}, we conclude that $P' \models \mathtt{go(Index)} \Rightarrow P \models false$.  

Note that linearisation required partial evaluation of the \texttt{perm} predicate, giving a blow-up in program size related to the length of the clause bodies.  This is further discussed at the end of Section \ref{experiments}.


\subsection{Obtaining linear over-approximations with a partial model}
\label{newton}

First we note that the set of predicates in $P^{\atmost{k}}$ is a subset of the set of predicates in $P^{\atmost{k+1}}$.
Given a model $M$ for the predicates in $P^{\atmost{k}}$,  $P^{\atmost{k+1}}$ can be linearised if we replace each occurrence of a predicate from $P^{\atmost{k}}$ in the body of a clause in $P^{\atmost{k+1}}$ with the corresponding constraint from the model $M$. 
The resulting set of clauses is linear since $P^{\atmost{k+1}}$ contains at most one predicate in its body from $P^{\atmost{k+1}}$ which is not in $P^{\atmost{k}}$.  Furthermore if $P^{\atmost{k+1}}$ has a model then so does the set of clauses resulting from the replacement;  the converse is however not the case since the  model $M$  represents an over-approximation of $P^{\atmost{k}}$. An example is given in  Section \ref{procverfication}.

More generally, we can replace any subset of the occurrences of predicates from $P^{\atmost{k}}$ in $P^{\atmost{k+1}}$. We summarise this in the following lemma.
\begin{lemma}[Linear over-approximation]
\label{linear}
Let $M$ be a model of the predicates in $P^{\atmost{k}}$, represented by a set of ``constrained facts" $p(X) \leftarrow \C$ where $p$ is a predicate in $P^{\atmost{k}}$.  Let $P'$ be any set of clauses obtained from $P^{\atmost{k+1}}$ by replacing some of the occurrences of predicates $p(X)$ from $P^{\atmost{k}}$ in the bodies of clauses in $P^{\atmost{k+1}}$ with their corresponding interpretation $\C$ in $M$. Then 
\begin{enumerate}
\item
If $P^{\atmost{k+1}}$ has a model then so does $P'$;
\item
If $P'$ contains no predicate from $P^{\atmost{k}}$, then $P'$ is linear.
\end{enumerate}
\end{lemma}

\section{Algorithm for solving sets of  non-linear Horn clauses}
\label{procverfication}
A basic
procedure for solving a set of non-linear Horn clauses using a linear Horn clause solver is presented in Algorithms \ref{alg:verify1} and \ref{alg2:verify2}. We use the term ``linear solver" for linear Horn clause solver for brevity. The main procedure \ProcSty{SOLVE(}\ArgSty{P}\ProcSty{)}  takes a set of non-linear Horn clauses $P$ as input and outputs (upon termination)  \emph{(safe, solution)} if \ArgSty{P} is solvable  or \emph{(unsafe, counterexample)} otherwise. We represent counterexample as a \emph{trace tree}. For a linear program it corresponds to a sequence of clauses used to derive a counterexample. 

\begin{definition}{$(S_{|t})$}
\label{cexprojection}
Let $S$ be an interpretation of a set of Horn clauses $P$. Let $t$ be any trace tree for some atom $A$ in $P$ (Definition \ref{def:tracetree}) and let $A_{t}$ be the set of heads of clauses with identifiers in $t$.  Then $S_{|t}$ is defined to be the set
$$S_{|t}=\{ (H \leftarrow \C) \mid  (H \leftarrow \C) \in S \wedge H \not\in A_{t}\}.$$


\end{definition}
Informally, the derivation corresponding to $t$ does not use any predicate interpreted by $S_{|t}$.  This notion is used in Algorithm \ref{alg2:verify2}.

Algorithm \ref{alg2:verify2} is an extended version of Algorithm \ref{alg:verify1}, which uses the solution for $P^{\atmost{k}}$ to help to linearise $P^{\atmost{k+1}}$ and also allows a more refined termination condition based on whether or not the solution for $P^{\atmost{k}}$ is used in constructing a counterexample for $P^{\atmost{k+1}}$.

The procedures make use of several sub-procedures which will be described next.

\subsection{Components of the algorithm}

\begin{itemize}
\item

\ProcSty{KDIM(}\ArgSty{P,k}\ProcSty{)}:  produces an at-most-$k$-dimension program 
$P^{\atmost{k}}$ (Definition \ref{kdim-most}). By definition, $P^{\atmost{k}}$ is linear for $k=0$. For our example program presented in Figure \ref{exprogram}, $Fib^{\atmost{0}}$ is shown in Figure \ref{fib0dim} which is linear since there is at-most one non-constraint atom in the body of each clauses.

\item
  \ProcSty{SOLVE\_LINEAR(}\ArgSty{P$^{\prime}$}\ProcSty{)}: solves a set of linear Horn clauses $P'$. We assume the following about a linear  solver: (i) if it terminates on $P'$, then it  returns either \emph{safe} and a \emph{solution} or \emph{unsafe} and a \emph{counterexample};  (ii) it is  sound, that is,  if it returns a \emph{solution $S$} for $P'$  then $P'$ has a model and $S$ is a solution (model) of $P'$; if it returns \emph{unsafe} and a counterexample \textit{cEx} then $P'$ is  unsafe and \textit{cEx} is a witness. In our setting (Algorithms \ref{alg:verify1} and \ref{alg2:verify2}), $P'$ corresponds to a linearised version of $P^{\atmost{k}}$ for some $P$ and $k\geq 0$. For technical reasons, the top level predicate $\mathit{false^{\exactly{k}}}$ of  $P^{\atmost{k}}$ if any, is renamed to $\mathit{false}$ before passing to a linear solver.

 In essence, any Horn clause solver which complies with our assumption, for example QARMC \cite{DBLP:conf/pldi/GrebenshchikovLPR12},  Convex polyhedral analyser \cite{kafleG2015horn}, ELDARICA \cite{DBLP:conf/fm/HojjatKGIKR12} etc. can be used in a black-box fashion but in this paper, we make use of a  solver described in \cite{kafleG2015horn}, which is based on abstract interpretation \cite{DBLP:conf/popl/CousotC77} over the domain of convex polyhedra \cite{DBLP:conf/popl/CousotH78} but  without refinement using finite tree automata. 
The solver produces the following solution for the program in Figure \ref{fib0dim}. We can check it is in fact a solution (model).

\begin{verbatim}
fib(0)(A,B) :- [-A>= -1,A>=0,B=A].
fib[0](A,B) :- [-A>= -1,A>=0,B=A].
false[0] :- <>. % <>  means that there is no model for false[0], 
			%so we can discard it
\end{verbatim}

\item
\ProcSty{LINEARISE(}\ArgSty{$P$,$k$,$S$}\ProcSty{)}  generates a linear set of clauses from $P^{\atmost{k}}$ and an interpretation $S$ for $P^{\atmost{k}}$. 
 Let $S$ be a set of constrained facts of the form $p(X) \leftarrow \C$, where $p$ is a predicate from $P^{\atmost{k}}$, the procedure replaces every clause from $P^{\atmost{k}}$ with head $p(X)$ by $p(X) \leftarrow \C$.  This produces a  set of clauses say $P'$.  Then the procedure  \ProcSty{LINEARISE\_PE(}\ArgSty{$P'$,$Index$}\ProcSty{)} is called, which is the linearisation procedure based on partial evaluation described in Section \ref{linearisation} where \ArgSty{$
Index$} is a bound for the stack usage for linearising $P^{\atmost{k}}$.
 

 
\end{itemize}

An excerpt from $Fib^{\atmost{1}}$ is shown below.

\begin{verbatim}
false(1) :- A>5, B<A,  fib(1)(A,B).
fib(1)(A,B) :- A>1, C=A-2, E=A-1, B=F+D,  fib(1)(C,D),  fib[0](E,F).
fib(0)(A,B) :- B=A,  A=<1, A>=0.
\end{verbatim}

After reusing  the solution  obtained for $Fib^{\atmost{0}}$ and linearising, we obtain the following set of  linear clauses. 
\begin{verbatim}
false(1) :-  A>5, B<A, fib(1)(A,B).
fib(1)(A,B) :-  -A>= -2, A>1, A-C=2,  B-D=1, fib(1)(C,D).
\end{verbatim}

Continuing to run our algorithm, the following solution obtained for $Fib^{\atmost{2}}$  becomes a solution for the program in Figure \ref{exprogram} (the original program) and the algorithm terminates. 

\begin{verbatim}

fib(0)(A,B) :- [-A>= -1,A>=0,B=1].
fib[0](A,B) :- [-A>= -1,A>=0,B=1].
fib(1)(A,B) :- [A>=2,A+ -B=0].
fib[1](A,B) :- [A+ -B>= -1,B>=1,-A+B>=0].
fib(2)(A,B) :- [A>=4,-2*A+B>= -3].
fib[2](A,B) :- [A>=0,B>=1,-A+B>=0].

\end{verbatim}

\begin{algorithm}{Procedure \ProcSty{SOLVE(}\ArgSty{P}\ProcSty{)}}
\caption{Algorithm for solving a set of Horn clauses}
\label{alg:verify1}
\SetKw{KwGoTo}{goto}
\KwIn{A set of CHCs $P$}
\KwOut{\emph{(safe, solution),  (unsafe, cex)}}
  $k \gets 0$\;
  $P' \gets \ProcSty{LINEARISE}(P,k,\emptyset)$\label{loop}\;
  $(\mathit{status}, \mathit{Result}) \gets \ProcSty{SOLVE\_LINEAR(}\ArgSty{P}{}^\prime\ProcSty{)}$ \tcc*[r]{Result is a solution or a cex}
  \uIf{$\mathit{status} =$ safe} {
       \lIf {($\mathit{Result}^{ \uparrow P^{\atmost{k}}}$  \text{ is a  solution  of } $P$)} { 
         \Return \( (\mathit{safe},\mathit{Result}^{ \uparrow P^{\atmost{k}}})\)
       }\; 
       $k \gets k+1$\; 
   } \Else{
      \Return \( (\mathit{unsafe}, \mathit{Result}) \) \tcc*[r]{Result  is a cex}
        }
   \KwGoTo \ref{loop}\;
\end{algorithm}

\begin{algorithm}{Procedure \ProcSty{SOLVE(}\ArgSty{P}\ProcSty{)}}
\caption{Algorithm for solving a set of Horn clauses, with reuse of lower dimension solutions}
\label{alg2:verify2}
\SetKw{KwGoTo}{goto}
\KwIn{A set of CHCs $P$}
\KwOut{\emph{(safe, solution),  (unsafe, cex)}}
  $k \gets 0$\;
  $S \gets \emptyset$\;
  $P' \gets \ProcSty{LINEARISE}(P,k,S)$\label{loop1}\;
  $(\mathit{status}, \mathit{Result}) \gets \ProcSty{SOLVE\_LINEAR(}\ArgSty{P}{}^\prime\ProcSty{)}$ \tcc*[r]{Result is a solution or a cex}
  \uIf{$\mathit{status} =$ safe} {
       \lIf {($\mathit{Result}^{ \uparrow P^{\atmost{k}}}$  \text{ is a  solution  of } $P$)} { 
         \Return \( (\mathit{safe},\mathit{Result}^{ \uparrow P^{\atmost{k}}})\)
       }\; 
       $k \gets k+1$\; 
       $S \gets \mathit{Result}$\;
   } \Else{
   \lIf(\tcc*[r]{Result  is a linear cex}){\(S = S_{| \mathit{Result} }\)}{\Return \( (\mathit{unsafe}, \mathit{Result})\)}

$S \gets S_{| \mathit{Result}}$\tcc*[r]{$S_{| \mathit{Result}}$: Definition \ref{cexprojection}}
   }
   \KwGoTo \ref{loop1}\;
\end{algorithm}

\begin{algorithm}{Procedure \ProcSty{LINEARISE(}\ArgSty{P, k, S}\ProcSty{)}}
\label{alg:linearise}
\caption{Algorithm for linearising a set of clauses}
\SetKw{KwGoTo}{goto}
\KwIn{A set of CHCs $P$, an integer $k$ and a set of constrained facts $S$}
\KwOut{A linearised set of clauses $P_{lin}$}
$P^{\atmost{k}} \gets  \ProcSty{KDIM(}\ArgSty{P,k}\ProcSty{)}$ \tcc*[r]{Definition \ref{kdim-most}}
$P' \gets$ \ProcSty{SUBSTITUTE(}\ArgSty{\(P^{\atmost{k}},S\)}\ProcSty{)}\tcc*[r]{substitute atoms of $P^{\atmost{k}}$ with their interpretations from $S$}
$Index \gets (i-1)*k+1$ \tcc*[r]{where $i$ is the maximal number of body atoms of $P$}
$P_{lin} \gets \ProcSty{LINEARISE\_PE(}\ArgSty{P}{}^\prime, \ArgSty{Index}\ProcSty{)}$ \tcc*[r]{Section \ref{pe}}
  \Return $P_{lin}$
\end{algorithm}

 \subsection{Reuse of solutions, refinement and linearisation} 
 Algorithm \ref{alg2:verify2} solves non-linear Horn clauses $P$ in essentially the same way as Algorithm \ref{alg:verify1}, but incorporates a \emph{refinement} phase in the case that the linear solver finds a counterexample. This counterexample possibly uses some of the model of the lower-dimension predicates $S$, in which case it is not certain whether it is a false alarm or a real counterexample.  If the counterexample did use some of the predicate solutions from $S$, then we discard those solutions (Algorithm \ref{alg2:verify2}, line 12) and return to the linearisation step.  If the counterexample does not use any predicate solutions from $S$, then it is a real counterexample (Algorithm \ref{alg2:verify2}, line 12). We will clarify this with an example program (linear for simplicity) shown below. 
 
 \begin{BVerbatim}
 c1. false:- X=0, p(X).
 c2. false:- q(X).
 c3. p(X):- X>0.
 c4. q(X):-X=0.
 \end{BVerbatim}
%
 
 Suppose we have an approximate solution \texttt{$S=\{ p(X):-true\}$} for the predicate $p(X)$. Using this solution, the above program is transformed into  the following  program.
 
  \begin{BVerbatim}
 c1. false:- X=0, p(X).
 c2. false:- q(X).
 c3. p(X):- true. (approximate solution)
 c4. q(X):-X=0.
 \end{BVerbatim}
 
The trace $c_1(c_3)$ is a counterexample for this transformed program  but not  to the original program (since it uses approximate solution for the predicate $p$). However  the trace $c_2(c_4)$ is a counterexample for this program as well as to the original  since it  does not use any approximate solution for the predicates appearing in the counterexample.
 
A schematic overview of Algorithm \ref{alg2:verify2} is shown in Figure \ref{fig:toolchain}. At each iteration of the \emph{abstraction-refinement} loop, the at-most-\(k\)-dimension under-approximation of $P$ is computed, then linearised and solved using a solver for linear Horn clauses.
 
\begin{figure}[H]
\centering
\adjustbox{max width=\linewidth}{
\begin{tikzpicture}
\draw[thick,  color=blue] (0.5,0) rectangle (13,6); 
\begin{scope}
 \node at (6,5.1) {\it CA -- Counterexample Analyser  };
\node at (3.5,5.6) {\it Lin -- Linearisation procedure };
\node at (9.2,5.6) {\it LS -- Linear Horn clause solver};

\draw[thick,dashed,green] (.6,1.2) rectangle (7.8,4.5);  
 \node at (4.2,4.1) {\bf   Abstraction };

\draw[thick,dashed,red] (8,1.2) rectangle (12.5,4.5);  
 \node at (9.8,4.1) {\bf   Refinement};


\node at (1.6,3.5) {CHC $P$}; 
\node at (2.8,3.0) { $k=0, S=\emptyset$}; 
 \draw[thick][->] (1.6,3.3) -- (1.6,2.5);

\draw[blue] (0.8,1.5) rectangle (2.0,2.5); 
\node at (1.5,2) {\emph{Lin} }; 
\node at (4.2,2.4) { $P', S,k$}; 
 \draw[thick][->] (2.0,2) -- (5.5,2);
 
\draw[blue] (5.5,1.5) rectangle (6.5,2.5); 
\node at (5.95,2.1) {\emph{LS}};
\node at (6.1,3.4) {({\bf safe, $R ^{\uparrow P^{\atmost{k}}})$}};
\node at (5.7,2.8) { $R$ };
\node at (7,2.8) {\emph{solution $P$?} };
\node at (6.7,1.2) {\emph{No} };
\draw[thick][->] (6.1,2.5) -- (6.1,3.2);
\node at (4.7,1.2) {$S \gets R, k=k+1$ };
\draw[thick][->] (6.1,1.5) -- (6.1,0.5);

 \node at (8.0,2.4) { $S, R,k$};
\draw[thick, blue][->] (6.5,2) -- (9.2,2);


\draw[blue] (9.2,1.5) rectangle (10.4,2.5); 
\node at (9.8,2.1) {\emph{CA}};

   \draw[thick][->] (1.5,0.5) -- (1.5,1.5);
   
 \draw[thick][->] (9.8,1.5) -- (9.8,0.5);
 \node at (11.4,1.0) {$S \gets  S_{ \mid R}, k$ };
  \draw[thick][-] (9.8,0.5) -- (1.5,0.5);
 
 \node at (9.8,3.4) { ({\bf unsafe, $R$})};
 \node at (11.4,2.8) {$S = S_{ | R} ?$};
\draw[thick][->] (9.8,2.5) -- (9.8,3.2);

\end{scope}
\end{tikzpicture}

}
\caption{\it Abstraction-refinement scheme for solving non-linear Horn clauses  using a solver for linear Horn clauses. $P'$ is a
set of linear CHC obtained by linearising the at-most-$k$-dimension underapproximation, \(P^{\atmost{k}}\), of \(P\).}
\label{fig:toolchain} 
\end{figure}

The soundness of  Algorithms \ref{alg:verify1} and \ref{alg2:verify2} is captured by Proposition \ref{soundness}.
\begin{proposition}[Soundness] 
\label{soundness}
  If Algorithm \ref{alg:verify1} or \ref{alg2:verify2} returns  safe and a solution $S$ for a set of clauses $P$ then $P$ is safe and $S$ is in fact a solution of $P$; if it returns unsafe and a counterexample cEx then $P$ is unsafe and cEx is a witness.
\end{proposition}

Another  property of the Algorithm \ref{alg2:verify2} is that of progress, that is, the same counterexample does not arise more than once. 



\section{Experimental results }
\label{experiments}

We made a prototype implementation   of  Algorithm \ref{alg2:verify2}  in the tool called \emph{LHornSolver}\footnote{\url{https://github.com/bishoksan/LHornSolver}}. It uses the solver described in \cite{kafleG2015horn} without refinement as a linear Horn clause solver.  \emph{LHornSolver} is written in Ciao prolog \cite{DBLP:journals/tplp/HermenegildoBCLMMP12} and is interfaced with Parma Polyhedra Library \cite{DBLP:journals/scp/BagnaraHZ08} and Yices SMT solver \cite{Dutertre:cav2014} for handling constraints. Then we carried out an experiment on a set of 44 non-linear CHC verification problems taken from  the repository\footnote{\url{https://svn.sosy-lab.org/software/sv-benchmarks/trunk/clauses/LIA/Eldarica/RECUR/}} of software verification benchmarks,  the recursive category of SV-COMP\footnote{\url{http://sv-comp.sosy-lab.org/2015/benchmarks.php}} \cite{DBLP:conf/tacas/000115} and  the tool QARMC. The experiments were run on a MAC computer running  OS X on  2.3 GHz Intel core i7 processor  and 8 GB memory. The benchmarks that we use in the experiments are not beyond the capabilities of existing solvers, but they are challenging. These programs are first translated to Prolog syntax  using the tools ELDARICA\footnote{\url{https://github.com/uuverifiers/eldarica}} \cite{DBLP:conf/fm/HojjatKGIKR12} and SeaHorn \cite{DBLP:conf/tacas/GurfinkelKN15}. Our aim with these  experiments  is to explore: (1) whether using a linear solver for non-linear problem solving  is practical;   (2) the relationship between the solvability of a problem and its dimension; and (3) how the current results compare with the results using the state of the art non-linear Horn clause verification tool (in our case RAHFT \cite{kafleG2015horn}). The results are summarized in Table~\ref{tbl:exp}.

  \begin{table}
    \begin{tabular}{|l|l|l|l|l|l|l|}
    \hline
                                                            & \textbf{RAHFT}    & ~        &  ~        & \textbf{LHornSolver} & ~       & ~         \\ \hline  \hline
    \textbf{Program}                                                 & \textbf{Safety}   & \textbf{\# iter.} & \textbf{Time (s)} & \textbf{Safety}       & \textbf{\#iter.} & \textbf{Time (s)} \\ \hline
    Addition03\_false-unreach                 &   safe   & 2        & $< 1$  & ?            &  ?      & ?        \\ \hline
     McCarthy91\_false-unreach &   unsafe & 0        & $< 1$  & ?            &  ?      & ?        \\ \hline
     addition.nts.pl                                        &   safe   & 0        & $< 1$  &  safe        & 1       & $<1$  \\ \hline
     bfprt.nts.pl                                           &   safe   & 0        & $< 1$  &  safe        & 2       & 4        \\ \hline
     binarysearch.nts.pl                                    &   safe   & 0        & $< 1$  &  safe        & 1       & 1.1      \\ \hline
     countZero.nts.pl                                       &   safe   & 0        & $< 1$  &  safe        & 1       & $< 1$  \\ \hline
     eq.horn                                                &   unsafe & 0        & $< 1$  &  unsafe      & 2       & $< 1$  \\ \hline
     fib.pl                                                 &   safe   & 0        & $< 1$  & ?            &  ?      & ?        \\ \hline
     identity.nts.pl                                        &   safe   & 0        & $< 1$  &  safe        & 1       & $< 1$  \\ \hline
     merge.nts.pl                                           &   safe   & 0        & $< 1$  &  safe        & 1       & 1.7      \\ \hline
     palindrome.nts.pl                                      &   safe   & 0        & $< 1$  &  safe        & 1       & $< 1$  \\ \hline
     parity.nts.pl                                          &   unsafe & 1        & $< 1$  & ?            &  ?      & ?        \\ \hline
     remainder.nts.pl                                       &   unsafe & 0        & $< 1$  &  unsafe      & 1       & $< 1$  \\ \hline
     revlen.pl                                              &   safe   & 0        & $< 1$  &  safe        & 1       & $< 1$  \\ \hline
     running.nts.pl                                         &   unsafe & 1        & $< 1$  & ?            &  ?      & ?        \\ \hline
     sum\_10x0\_false-unreach                     &   unsafe & 10       & 10       & ?            &  ?      & ?        \\ \hline
     sum\_non\_eq\_false-unreach                  &   unsafe & 0        & $< 1$  & ?            &  ?      & ?        \\ \hline
     suma1.horn                                             &   unsafe & 0        & $< 1$  &  unsafe      & 1       & $< 1$  \\ \hline
     suma2.horn                                             &   unsafe & 0        & $< 1$  &  unsafe      & 2       & $< 1$  \\ \hline
     summ\_SG1.r.horn                                       &   safe   & 0        & $< 1$  & ?            &  ?      & ?        \\ \hline
     summ\_SG2.r.horn                                       &   safe   & 8        & 78       & ?            &  ?      & ?        \\ \hline
     summ\_SG3.horn                                         &   safe   & 0        & $< 1$  &  safe        & 1       & $< 1$  \\ \hline
     summ\_b.horn                                           &   safe   & 2        & 1.7      & ?            &  ?      & ?        \\ \hline
     summ\_binsearch.horn                                   &   safe   & 1        & 3        & ?            &  ?      & ?        \\ \hline
     summ\_cil.casts.horn                                   &   safe   & 0        & $< 1$  &  safe        & 1       & $< 1$  \\ \hline
     summ\_formals.horn                                     &   safe   & 0        & $< 1$  &  safe        & 1       & $< 1$  \\ \hline
     summ\_g.horn                                           &   safe   & 0        & $< 1$  & ?            &  ?      & ?        \\ \hline
     summ\_globals.horn                                     &   safe   & 0        & $< 1$  &  safe        & 1       & $< 1$  \\ \hline
     summ\_h.horn                                           &   safe   & 0        & $< 1$  &  safe        & 2       & $< 1$  \\ \hline
     summ\_local-ctx-call.horn                              &   safe   & 0        & $< 1$  &  safe        & 1       & $< 1$  \\ \hline
     summ\_locals.horn                                      &   safe   & 0        & $< 1$  & ?            &  ?      & ?        \\ \hline
     summ\_locals2.horn                                     &   safe   & 0        & $< 1$  &  safe        & 1       & $< 1$  \\ \hline
     summ\_locals3.horn                                     &   safe   & 0        & $< 1$  &  safe        & 1       & $< 1$  \\ \hline
     summ\_locals4.horn                                     &   safe   & 0        & $< 1$  &  safe        & 2       & 2.2      \\ \hline
     summ\_mccarthy2.horn                                   &   safe   & 3        & 5        & ?            &  ?      & ?        \\ \hline
     summ\_multi-call.horn                                  &   safe   & 0        & $< 1$  &  safe        & 1       & $< 1$  \\ \hline
     summ\_nested.horn                                      &   safe   & 0        & $< 1$  &  safe        & 1       & $< 1$  \\ \hline
     summ\_ptr\_assign.horn                                 &   safe   & 0        & $< 1$  &  safe        & 1       & $< 1$  \\ \hline
     summ\_recursive.horn                                   &   safe   & 0        & $< 1$  & ?            &  ?      & ?        \\ \hline
     summ\_rholocal.horn                                    &   safe   & 0        & $< 1$  &  safe        & 1       & $< 1$  \\ \hline
     summ\_rholocal2.horn                                   &   safe   & 0        & $< 1$  &  safe        & 1       & $< 1$  \\ \hline
     summ\_slicing.horn                                     &   safe   & 0        & $< 1$  & ?            &  ?      & ?        \\ \hline
     summ\_summs.horn                                       &   safe   & 0        & $< 1$  & ?            &  ?      & ?        \\ \hline
     summ\_typedef.horn                                     &   safe   & 0        & $< 1$  &  safe        & 1       & $< 1$  \\ \hline
     summ\_x.horn                                           &   safe   & 0        & $< 1$  & ?            &  ?      & ?        \\ \hline \hline
     Average                                                    & ~  &    0.64 & 2.3 & ~ & 1.185 & $< 1$   \\ \hline
    \end{tabular}
    \caption{Experimental  results on non-linear CHC verification problems with a timeout of 5 minutes.}
    \label{tbl:exp}
\end{table}
 
In the table {\it Program} represents a program, {\it Safety} represents a verification result, {\it \#iter.}  and  {\it Time (s)} successively represent the number of refinement iterations and the time in seconds need to solve a program using both RAHFT and LHornSolver. It is to note that the underlying abstract interpreter, that is, the convex polyhedral analyser (CPA)  is the same for both \emph{RAHFT} and \emph{LHornSolver} but \emph{LHornSolver} uses it to solve linear Horn clauses though the CPA is not optimised for linear problems. The column {\it \#iter.} for LHornSolver represents a value of $k$ for which a solution of $P^{\atmost{k}}$ (under-approximation) of a set of clauses $P$ becomes a solution for $P$ or $P^{\atmost{k}}$ becomes unsafe. The symbol ``?'' means that the result is unknown within the given time bound. The result ``safe'' means that the program is safe (solvable) and ``unsafe'' means it is unsafe. 

\emph{LHornSolver} solves 27 out of 44 (about $61\%$) problems within a second.
In most of these problems, a solution of an under approximation ($P^{\atmost{k}}$) becomes a solution for the original program or $P^{\atmost{k}}$ becomes unsafe for a fairly small value of $k$ ($1$ or $2$). This suggests that the solvability of a problem is shallow with respect to its dimension. This demonstrates the feasibility of solving a set of non-linear Horn clauses  using a  solver for linear Horn clauses. 

In contrast, RAHFT solves all the problem. The difference in results maybe due to the following reason: 
the linear solver that is used in LHornSolver is the CPA (without refinement in contrast to \cite{kafleG2015horn}). The solver terminates but produces \emph{false alarms}.  If we use CPA with refinement as in \cite{kafleG2015horn}, then we lose predicates names (due to program transformation), so the solution or counterexamples produced by the tool do not correspond to the original program (it is very hard to keep track of the changes). This hinders  the reuse of solution from lower dimension to linearise program of higher dimension or refine it using the counterexample trace. Other solvers which don't modify the programs but produce solutions or counterexamples can be used as a linear solver in principle and we leave it for the future work.   Another disadvantage of  using  CPA  is that, if it cannot solve a linear program, then it emits an abstract trace which is checked for a feasibility. If it is spurious then LHornSolver returns with \emph{unknown} (in principle we can refine the program but the refinement will have the problem as mentioned above). So it is highly unlikely that the trace picked by the tool non-deterministically results to be a real counterexample. We noticed in our experiments that the trace picked was spurious most of the times and \emph{LHornSolver} immediately returned ``unknown'' answer before the timeout. This also explains why solving time of \emph{LHornSolver} is less than that of \emph{RAHFT}.


The interpreter described in Figure \ref{fig:interpreter} computed a \emph{permutation} of the atoms in a clause body; partial evaluation of the permutation procedure can cause a blow-up of the size of the linearised program, relative to the number of atoms in clause bodies. During our experiment we found that the maximum number of atoms in the bodies  of the clauses in our benchmark programs was 5 and the value of $k$ was relatively small ($k=0 \ldots 2$). The permutation procedure can be avoided if we first generate an at-most-$k$-dimension program whose body atoms are ordered by increasing dimension. This needs unfolding of the $\epsilon$-clauses, since atoms whose predicate is $p^{\atmost{d}}$ cannot be ordered directly;  only atoms with predicates of the form $p^{\exactly{d}}$can be ordered. We have not yet evaluated the trade-offs in these two approaches.

\section{Related Work}
\label{rel}

In the world of Horn clause solvers, after fixing a constraint theory, we can
distinguish solvers depending on whether they can handle general non-linear Horn
clauses or not. A majority of solvers
\cite{DBLP:conf/tacas/GurfinkelKN15,DBLP:conf/tacas/GrebenshchikovGLPR12,DBLP:conf/cav/RummerHK13,McmillanR2013,kafleG2015horn}
handle non-linear Horn clauses but there are notable exceptions like
VeriMAP~\cite{DBLP:conf/tacas/AngelisFPP14} or
Sally\footnote{\url{https://github.com/SRI-CSL/sally}}. For both VeriMAP and Sally, their
underlying reasoning engine handles only linear Horn clauses which restricts,
in principle, their applicability.  Our contribution is to lift this
restriction by allowing those tools to be applied on arbitrary sets of
Horn Clauses, linear or not, through a linearisation procedure that
underapproximates the set of solutions.  We give empirical evidence that this
underapproximation often provides enough coverage to enable the verification
of the original set of Horn clauses. To summarize, we allow solvers with restrictions
to be applied on any input at the price of an under-approximation which often results in full coverage.


Our linearisation method  based on partial evaluation described in Section \ref{pe} is related to the linearisation method 
based on \emph{fold-unfold transformations} described by De Angelis \emph{et al.} \cite{DBLP:journals/tplp/AngelisFPP15}. While their procedure transforms the target set of clauses directly, we transform an interpreter for the clauses using a generic partial evaluation procedure.  Any clause transformation procedure could be formulated as a meta-program and partial evaluation applied to that program to yield the specified transformation. Thus neither approach offers any more power than the other.  However the use of partial evaluation is arguably more flexible. The interpreter that is partially evaluated in our procedure is a standard interpreter for Horn clauses, modified with a bound on the size of goals, directly incorporating a general result that there is an upper bound on the size of goals 
in derivations with dimension-bounded programs. This provides a very generic starting point for the transformation with an explicit relation to the semantics of the clauses. A whole family of similar transformations could be formulated by varying the interpreter (for example using breadth-first search).  The procedure in \cite{DBLP:journals/tplp/AngelisFPP15} is tailored to a
restrictive setting where only goal clauses (integrity constraints) are non-linear and rest of the clauses are linear;  correctness has to be established for that case.

Ganty, Iosif and Kone\v{c}n\'{y}~\cite{DBLP:conf/tacas/GantyIK13} used the notion of
\emph{tree dimension} for computing summaries of procedural programs by
underapproximating them. Roughly speaking, they compute procedure summaries
iteratively, starting from the program behaviors captured by derivation trees
of dimension \(0\). Then they reuse these summaries to compute summaries for
program behaviors captured by derivation trees of dimension \(1\) and so on for
\(2\), \(3\), etc. Kafle, Gallagher and Ganty
\cite{DBLP:journals/corr/KafleGG15} adapted the idea of dimension-based
underapproximations to the setting of Horn clause systems. They gave empirical
evidence supporting the thesis that for small values of the dimension the
solutions are general enough to hold for every dimension. Their approach still
required the use of general Horn-clause solvers capable of handling non-linear
clauses. In this paper, we lift this requirement and allow the use of solvers
for linear clauses only. Moreover, we provide an abstraction refinement loop
that enables the solutions for lower dimension to be reused when searching for
solutions in higher dimension.


\section{Conclusion and future work}

We presented an \emph{abstraction-refinement} approach for solving a set of non-linear Horn clauses using an off-the-shelf linear Horn clause solver.  It was achieved through a linearisation of a dimension bounded set of Horn clauses (which are known to be linearisable) using partial evaluation  and the use of a linear Horn clause solver.  Experiment  on a set of non-linear Horn clause verification problems using our approach shows that the approach is feasible (a linear solver can be used for solving non-linear problems) and the solvability of a problem is shallow with respect to its dimension. 

A linear set of clauses is essentially a transition system.
Many tools exist whose input languages have a form such as C programs (without procedure calls), control flow graphs, Boogie programs, and such formalisms whose semantics is usually given as a transitions system. The results of this paper suggest that such tools could be applied to the verification of non-linear Horn clauses.  

In the future, we plan to compare our results with the results from a specialised linear Horn clause solver like VeriMap and other non-linear Horn clause solvers. We also plan to experiment with  different linearisation strategies for Horn clauses and study their effects in Horn clause verification.

%

\section*{Acknowledgement}
The authors would like to thank Jos\'{e} F. Morales for his help with Ciao Prolog foreign language interface and some parts of the implementation.

\bibliographystyle{eptcs}
\bibliography{refs}
\end{document}